\RequirePackage{lineno}
\documentclass[traditabstract,twocolumn]{aa}

\usepackage{amsmath}
\usepackage{amssymb}
\usepackage{newcent}
\usepackage{txfonts}
\usepackage{graphicx}
\usepackage{natbib}
\usepackage{subfigure}
\usepackage{color}

\bibpunct{(}{)}{;}{a}{}{,} 

\newcommand{\Fermic}{\textit{Fermi}}
\newcommand{\Fermi}{\Fermic\ }
\newcommand{\FermiLATc}{\Fermic\ LAT}
\newcommand{\FermiLAT}{\FermiLATc\ }

\newcommand{\hmsh}{\mbox{$^{\mathrm h}$}}%
\newcommand{\hmsm}{\mbox{$^{\mathrm m}$}}%
\newcommand{\hmss}{\mbox{$^{\mathrm s}$}}%
\newcommand{\HMS}[3]{$#1\hmsh\,#2\hmsm\,#3\hmss$}
\newcommand{\DMS}[3]{$#1\arcdeg\,#2\arcmin\,#3\arcsec$}
\newcommand\arcdeg{\mbox{$^\circ$}}%

\newcommand{\NTeVCatAGN}{50 }
\newcommand{\NTeVCatBLLac}{37 }

\newcommand{\NTeVCatDefaultcatalog}{30 }
\newcommand{\NTwoLACTeVBlazar}{36 }

\newcommand{\NTeVCatDefaultCatalogFSRQ}{two }

\newcommand{\NSamplewoRG}{23 }
\newcommand{\half}{{\frac{1}{2}}}

\begin{document}
\title{Evidence for a cosmological effect in $\gamma$-ray spectra of BL~Lacs.}

\author{D.A.~Sanchez\inst{1} \and S.~Fegan \inst{2}
	\and  B.~Giebels\inst{2}}

\institute{Max-Planck-Institut f\"ur Kernphysik, P.O. Box 103980, D 69029 Heidelberg, Germany
	\and Laboratoire Leprince-Ringuet, Ecole polytechnique, CNRS/IN2P3, F-91128 Palaiseau, France}

\date{Received ; Accepted}

\abstract {We update the list of GeV-TeV extragalactic $\gamma$-ray sources using the 2-year catalog from the \FermiLAT and recent results ground-based $\gamma$-ray telescopes. Breaks in the spectra between the high energy (100\,MeV\,$<E<$\,300\,GeV) and the very high energy (E$>$\,200\,GeV) ranges, and their dependence on distance, are discussed in the context of absorption on the extragalactic background light (EBL). We calculate the size of the expected break using a model for the EBL and compare it to the data taking into account systematic uncertainties in the measurements. We develop a novel Bayeasian model to describe this dataset and use it to constrain two simple models for the EBL-induced breaks.}

\titlerunning{Evidence for a cosmological effect in $\gamma$-ray spectra of BL~Lacs.}

\maketitle

\section{Introduction}

The extragalactic background light (EBL) is a diffuse field of U.V., optical and infra-red photons, with wavelengths in the range $\lambda=0.1\,{\rm -} 1000 \mu{\rm m}$, on which the integrated history of star formation in the Universe is imprinted. The spectral energy distribution of the EBL consists of two distinct components: the first, peaking in $\nu F_\nu$ around $\simeq 1 \mu$m and commonly referred to as the cosmic optical background (COB), was produced by thermal emission from stars since the big bang. The second component, peaking at longer wavelengths ($\simeq 100 \mu$m), having comparable peak energy density to the COB and being referred to as the cosmic infra-red background (CIB), originates from the absorption and reemission of starlight by dust \citep[see][for review]{EBL_REVIEW}.

Direct measurements of the EBL density are difficult due to local foregrounds, such as the zodiacal light and Galactic radiation \citep{EBL_REVIEW}, and are often interpreted as upper limits, while galaxy number counts in optical or infrared provide lower limits \citep{REF::GALAXY_COUNTS}.

Since $\gamma$ rays of observed energy $E_\gamma$ can interact with EBL photons of energy $E_{\rm EBL}$ at a redshift $z$ through $\gamma \gamma \rightarrow e^+ + e^-$ when $E_\gamma/1\,{\rm TeV}>0.26\,{\rm eV}/E_{\rm EBL}(1+z)$, the spectra of distant extragalactic sources measured in the very high energy (VHE, E$>$\,200\,GeV) regime should differ from their emitted (intrinsic) spectra if the EBL density is nonzero\footnote{The created pairs can also upscatter CMBR photons to high energy $\gamma$-rays \citep{1986MNRAS.221..769P} and induce yet another spectral distortion mostly at energies $\sim 100\,{\rm GeV}$ and below (e.g., \citealt{2002A&A...384..834A,2007A&A...469..857D}), a feature which has currently not <been observed \citep{2010Sci...328...73N} with instruments sensitive in that energy range.}. Since a large fraction of the emitted power in BL~Lac-type blazars is in $\gamma$-ray band, this must be accounted for when spectral energy distributions are used to model their underlying physical properties \citep{1999APh....11...35C}.

Finding clear evidence for this EBL-induced attenuation has proven remarkably difficult to date. The fall-off in the EBL spectral density between the COB and CIB peaks (around $0.1\,{\rm eV}$) should be visible as a kink in the measured VHE spectra around $1\,{\rm TeV}$. This was sought for, e.g. in the blazar H~1426+428 by \citet{2003A&A...403..523A}, but results have been inconclusive given the large statistical errors. The signature of the EBL should also be evident in studies of the global population of VHE sources, since the energy-dependent attenuation increases with distance, such that the observed spectra are expected to become softer, i.e. the photon index, $\Gamma$, in power-law spectral fits should increase with redshift, $z$. Such studies have not been successful either, with no evidence for a redshift-dependent effect being found by \citet{2008arXiv0809.3625M},  \citet[][the authors attributing this to varying spectral states inducing a large scatter in the data]{MNL2:MNL20602} or \citet[][see in particular Figure 13]{2011ApJ...733...77O}.

It has however been possible to constrain the EBL density in the energy range where they interact with observed $\gamma$ rays. Using VHE spectra from distant BL~Lac objects and a theoretically motivated conjecture that the photon index of the intrinsic spectrum cannot be harder than $\Gamma_I\simeq 1.5$, several authors have derived upper limits for the density close to the lower limits from galaxy counts \citep{REF::KRENNRICH_DWEK_EBL,REF::HESS_EBL_Limite}. Recently, \citet{2012Sci...338.1190A} and \citet{2012arXiv1212.3409H} have measured the EBL density using its imprint in the spectra of BL~Lac objects and found a density of EBL compatible with the best upper limits to date  \citep{MRMH2012}.

Operating as an all-sky monitor, the \FermiLAT \citep{2009ApJ...697.1071A} observes $\gamma$ rays in the high energy (HE, 100\,MeV\,$<E<$\,300\,GeV) range, where the effects of the EBL are much smaller than in the VHE. Sources detected both in the HE by \Fermi and in the VHE by imaging atmospheric Cherenkov telescopes (IACTs) then provide an opportunity to probe the effects of the energy-dependent attenuation from the EBL absorption across a much wider energy range. Here we present an updated list of GeV-TeV sources, building on the work of \citet{2009ApJ...707.1310A} and \citet{2011arXiv1108.1420T}.

\section{Selection of the sources}
\label{selection}

Since the first detection of an extragalactic $\gamma$-ray source in the VHE range by Whipple \citep{TEV_MARKARIAN_421}, \NTeVCatAGN AGN have been discovered in this energy band. The rate of detections has increased dramatically with the increased sensitivity of the latest generation of IACTs (VERITAS, MAGIC and H.E.S.S.) and an improved observation strategy using data from the \Fermi-LAT. An up to date view of the VHE sky can be found by browsing the TeVCat catalogue\footnote{http://tevcat.uchicago.edu/} \citep{REF::TEVCAT}.

We select AGN from TeVCat for which a HE and VHE spectrum has been published and a firm redshift has been determined. Of the \NTeVCatAGN AGN, \NTeVCatDefaultcatalog are BL~Lacs with published spectral information. Three of these, namely 3C~66A, PKS~1424+240 and PG~1553+113, do not have a firm redshift determination. In addition there are \NTeVCatDefaultCatalogFSRQ VHE-detected FSRQs, 3C~279 and 4C~+21.35. Despite their large redshift (see Table \ref{table:list}), internal absorption close to the emission region may strongly affect their spectra \citep{INTRINSIC_ABSORPTION_AHARONIAN,3C279_INTER_ABS}, hence we include them in this study only for illustrative purposes. We also include the radio galaxies Centaurus~A and M~87 in the sample.

The second \Fermi catalogue of AGN \citep[2LAC,][]{2011arXiv1108.1420T} includes 1057 sources associated with AGN of many kinds. In this list, \NTwoLACTeVBlazar out of the \NTeVCatBLLac VHE BL~Lacs have a \Fermi counterpart, the six that are not detected being SHBL~J001355.9-185406, 1ES~0229+200, 1ES~0347-121, PKS~0548-322, HESS~J1943+213, 1ES~1312-423.

By merging the two lists, our sample contains \NSamplewoRG HE-VHE BL~Lacs and two radio galaxies. The characteristics of these sources are listed in Table~\ref{table:list}, including the photon indexes from published power-law spectral fits in the HE (from 2LAC) and VHE (from reference given in the table), which form the dataset for this study. 

AGN are observed to be variable in all wavelengths. In the VHE regime, flaring episodes have been observed from a number of such sources, in particular those that are bright and/or close by (such as Mrk~421, Mrk~501 and PKS~2155$-$304), but most have not shown clear evidence of variability - many have been detected close to the sensitivity threshold of the instrument. It is not improbable that variability is a common feature of HBLs in the VHE regime, future instruments will be able to probe this in a larger population of fainter, more distant sources. In the HE range, BL~Lacs, and especially HBL, are found to be the less variable \citep{2010ApJ...722..520A}. To reduce any bias introduced by the use of non-simultaneous observations, we use the VHE spectrum which has the lowest flux reported in the literature, resulting in a generally good agreement between the overlapping energy ranges \citep{2009ApJ...707.1310A}.

\begin{table*}[p]
\caption{List of HE-VHE BL Lacs and radio galaxies. Only statistical errors are given. } 
\label{table:list} 
\centering 
\begin{tabular}{c c c c c c c c} 
\hline \hline 
Source Name & $\alpha_{\rm J2000}$ & $\delta_{\rm J2000}$ & Type & $z$ &$\Gamma_{HE}$ &$\Gamma_{VHE}$ & Ref. \\
\hline 
Centaurus A   & \HMS{13}{25}{27.6} & \DMS{-43}{01}{09} & FR1  & $0.00183$ & $2.76\pm0.05$ & $2.7\pm0.5$  & [1] \\
M 87          & \HMS{12}{30}{49.4} & \DMS{+12}{23}{28} & FR1  & $0.004233$ & $2.17\pm0.07$ & $2.60\pm0.35$ & [1] \\
Markarian 421 & \HMS{11}{04}{27.3} & \DMS{+38}{12}{32} & HBL  & $0.031$  & $1.77\pm0.01$ & $2.20\pm0.08$   & [1] \\
Markarian 501 & \HMS{16}{53}{52.2} & \DMS{+39}{45}{37} & HBL  & $0.034$  & $1.74\pm0.03$ & $2.54\pm0.70$   & [1] \\
1ES 2344+514  & \HMS{23}{47}{04.8} & \DMS{+51}{42}{18} & HBL  & $0.044$  & $1.72\pm0.08$ & $2.95\pm0.12$   & [1] \\
Markarian 180 & \HMS{11}{36}{26.4} & \DMS{+70}{09}{27} & HBL  & $0.046$  & $1.74\pm0.08$ & $3.3\pm0.7$   & [1] \\
1ES 1959+650  & \HMS{19}{59}{59.9} & \DMS{+65}{08}{55} & HBL  & $0.048$  & $1.94\pm0.03$ & $2.58\pm0.18$   & [1] \\
BL Lacertae   & \HMS{22}{02}{43.3} & \DMS{+42}{16}{40} & LBL  & $0.069$  & $2.11\pm0.04$ & $3.6\pm0.6$   & [1] \\
PKS 2005-489  & \HMS{20}{09}{25.4} & \DMS{-48}{49}{54} & HBL  & $0.071$  & $1.78\pm0.05$ & $3.20\pm0.16$   & [8] \\
RGB J0152+017 & \HMS{01}{52}{39.6} & \DMS{+01}{47}{17} & HBL  & $0.080$  & $1.79\pm0.14$ & $2.95\pm0.36$   & [1] \\
W Comae       & \HMS{12}{21}{31.7} & \DMS{+28}{13}{59} & IBL  & $0.102$  & $2.02\pm0.03$ & $3.81\pm0.35$   & [1] \\
PKS 2155-304  & \HMS{21}{58}{52.1} & \DMS{-30}{13}{32} & HBL  & $0.117$  & $1.84\pm0.02$ & $3.32\pm0.06$   & [1] \\
B3 2247+381   & \HMS{22}{50}{06.6} & \DMS{+38}{25}{58} & HBL  & $0.119$  & $1.83\pm 0.11$ & $3.2\pm0.6$   & [9] \\
RGB J0710+591 & \HMS{07}{10}{30.1} & \DMS{+59}{08}{20} & HBL  & $0.125$  & $1.53\pm0.12$ & $2.69\pm0.26$   & [5]  \\
H 1426+428    & \HMS{14}{28}{32.7} & \DMS{+42}{40}{21} & HBL  & $0.129$  & $1.32\pm0.12$ & $3.5\pm0.35$   & [1] \\
1ES 0806+524  & \HMS{08}{09}{49.2} & \DMS{+52}{18}{58} & HBL  & $0.138$  & $1.94\pm0.06$ & $3.6\pm1.0$   & [1] \\
1RXS J101015.9-311909  & \HMS{10}{10}{15.03} & \DMS{-31}{18}{18.4} & HBL  & $0.142$ & $2.09\pm0.15$ & $3.08\pm.42$   & [7] \\
H 2356-309    & \HMS{23}{59}{07.9} & \DMS{-30}{37}{41} & HBL  & $0.167$  & $1.89\pm0.17$ & $3.06\pm0.15$   & [10] \\
RX J0648.7+1516& \HMS{06}{48}{45.6} & \DMS{+15}{16}{12} & HBL  & $0.179$ & $1.74\pm 0.11$ & $4.4\pm0.8$   & [4] \\
1ES 1218+304  & \HMS{12}{21}{21.9} & \DMS{+30}{10}{37} & HBL  & $0.182$  & $1.71\pm0.07$ & $3.08\pm0.34$   & [1] \\
1ES 1101-232  & \HMS{11}{03}{37.6} & \DMS{-23}{29}{30} & HBL  & $0.186$  & $1.80\pm0.21$ & $2.94\pm0.20$   & [1] \\
RBS 0413& \HMS{03}{19}{51.8} & \DMS{+18}{45}{34} & HBL  & $0.19$ & $1.55\pm 0.11 $ & $3.18\pm0.68$   & [2] \\
1ES 1011+496  & \HMS{10}{15}{04.1} & \DMS{+49}{26}{01} & HBL  & $0.212$  & $1.72\pm0.04$ & $4.0\pm0.5$   & [1] \\
1ES 0414+009& \HMS{04}{16}{52.4} & \DMS{+01}{05}{24} & HBL  & $0.287$ & $1.98\pm 0.16$ & $3.45\pm0.25$   & [3] \\
S5 0716+714   & \HMS{07}{21}{53.4} & \DMS{+71}{20}{36} & LBL  & $0.300$  & $2.00\pm0.02$ & $3.45\pm0.54$   & [6] \\
\hline \hline
\multicolumn{8}{l}{Additional sources used for illustration only:} \\
\hline 
3C 66A        & \HMS{02}{22}{39.6} & \DMS{+43}{02}{08} & IBL  & $0.444$?  & $1.85\pm0.02$ & $4.1\pm0.4$ & [1] \\
4C +21.35     & \HMS{12}{24}{54.4} & \DMS{+21}{22}{46} & FSRQ & $0.432$  & $ 1.95\pm 0.21$ & $3.75\pm0.27$ & [11] \\
PG 1553+113   & \HMS{15}{55}{43.0} & \DMS{+11}{11}{24} & HBL  & $0.43-0.58$    & $1.67\pm0.02$ & $4.41\pm0.14^\dagger$   & [12]  \\
3C 279        & \HMS{12}{56}{11.2} & \DMS{-05}{47}{22} & FSRQ & $0.536$  & $2.22\pm0.02$ & $4.1\pm0.7$ & [1] \\
\hline 
\end{tabular}
\tablebib{
[1]~\citet[][see references therein]{2009ApJ...707.1310A}; [2]~\citet{2012arXiv1204.0865V}; [3]~\citet{1es0414}; [4]~\citet{rxj0648}; [5]~\citet{rgbj0710}; [6]~\citet{S50716}; [7]~\citet{2012arXiv1204.1964H}; [8]~\citet{pks2005}; [9]~\citet{2012A&A...539A.118A}; [10]~\citet{h2356}; [11]~\citet{PKS1222+21}; [12]~\citet{2011arXiv1110.0038W}.\\
$\dagger$ \citet{2012ApJ...748...46A} recently published long-term observations on this object and derived a compatible photon index.
}
\end{table*}

\section{Interpretation}
\label{discussion}
\subsection{Spectral evolution with the redshift}
\label{pheno}
The mean HE and VHE indexes of our sample are $\langle\Gamma_{HE}\rangle = 1.86$ and $\langle\Gamma_{VHE}\rangle = 3.18$, respectively.  For each source the photon index measured in the VHE range is greater than or compatible with that found in the HE, i.e. $\Delta\Gamma = \Gamma_{\rm VHE}-\Gamma_{\rm HE}\gtrsim 0$. In the HE band, the RMS of the measured indexes is $\sigma_{HE}=0.26$, and the \textit{excess variance}, which accounts for the measurement errors\footnote{We define the excess variance of a set of measured quantities $x_i\pm\sigma_i$ as $(\sigma^{XS})^2 = \langle x_i^2 \rangle - \langle x_i\rangle^2 - \langle \sigma_i^2\rangle$.}, is $\sigma^{XS}_{HE}=0.24$. In the VHE regime, the RMS is $\sigma_{VHE}=0.49$, while the excess variance is $\sigma^{XS}_{VHE}=0.10$, showing that most of the sample variance can be ascribed to the errors on the individual measurements rather than to the intrinic distribution.

\begin{figure*}[p]
\centering
\includegraphics[width=0.99 \textwidth]{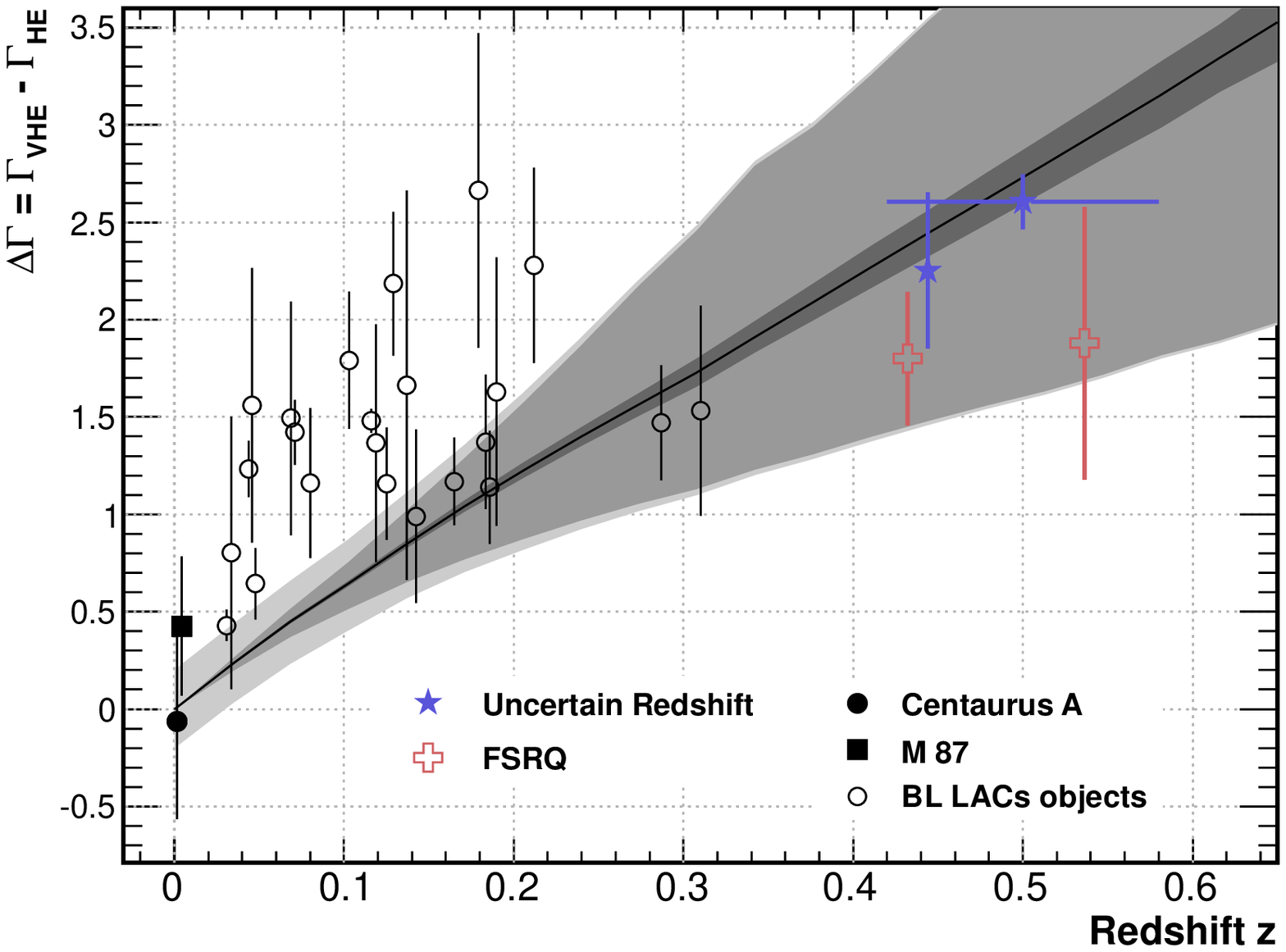}
\caption{The value of $\Delta \Gamma$ as a function of the redshift $z$. The black line is the theoretical break obtained with \citet{2008A&A...487..837F} model. The uncertainties due to the energy resolution of IACTs (dark gray), the different threshold energies (gray) and the systematic errors of 0.2 (light gray) are shown. The FSRQs and the BL~Lacs with uncertain redshift are shown for illustration.}
\label{figure}
\end{figure*}

The points on Figure~\ref{figure} show $\Delta \Gamma$ versus the redshift $z$ for our sample of sources. The two close-by radio galaxies do not show significant spectral breaks. For all other sources $\Delta\Gamma\gtrsim0.5$, and those more distant than $z=0.1$ exhibit a break of $\Delta\Gamma\gtrsim1.0$. A dependence of $\Delta\Gamma$ with $z$ is apparent in our sample.

If we were to assume that the intrinsic spectrum of each object was well represented by a single power law across the entire HE and VHE domain, as seems to be the case for the two nearby radio galaxies for which $\Delta\Gamma\sim0$, we should expect that any significant break in the measured spectrum is the result of absorption on the EBL, i.e.\ $\Gamma_{VHE} = \Gamma_{Int} + \Delta\Gamma_{EBL}(E,z) \approx \Gamma_{Int} + \frac{d\tau}{d\log E} (E,z)$, where $\tau(E_\gamma,z)$ is the optical depth due to the attenuation by pair production \cite[][Equation~2]{2009ApJ...707.1310A}. To leading order the expected EBL break increases linearly in the redshift and we would $\Delta\Gamma$ to be correlated with $z$. The Pearson correlation factor for our dataset is $\rho = 0.56\pm 0.11$, more than $5 \sigma$ away from 0, showing clear evidence that this dependency exists.

In appendix~\ref{App3} we outline a simple method to evaluate the size of the break that might be expected from redshifting a curved intrinsic SSC spectrum within the HE and VHE observation windows, which gives rise a K correction for the measured spectral indexes. This study shows that such a K correction would account for $\sim15$\% of the observed break for the most distant source in the sample at $z=0.3$.

\subsection{Expected EBL-induced spectral break}
\label{Prediction}

To further evaluate the data, we estimate the size of the spectral break as a function of redshift from the EBL density model of \citet[hereafter Fra08]{2008A&A...487..837F}. We perform a simple simulation in which a hypothetical source with a flat spectrum ($\Gamma_I=0$) is placed at a distance $z$ and its flux attenuated by the EBL. The simulated spectrum consists of 20 logarithmically spaced bins per decade, equally weighted, which is fitted with a power law model above 200 GeV to evaluate the measured index. The limited photon flux at the highest energies is accounted for in an ad hoc manner; the upper energy bound of the fit is chosen to be the point at which the differential flux is 1\% of the flux at 200\,GeV (up to a maximum of 10\,TeV).

The predicted $\Delta \Gamma (z)$ is the black line shown in Figure~\ref{figure}. The shaded gray areas show uncertainties on this calculation, which arise from:
\begin{itemize}
\item the $\sim10$\% energy resolution typical of IACTs which is taken into account by shifting the energy bins by $\pm 10$\% (dark gray area), 
\item the threshold energy of the observations, which can vary from 100 GeV to more than 500 GeV (gray area), and
\item the systematic error on the measured photon spectral index, typically 0.2 (light gray).
\end{itemize}

It is clear from Figure~\ref{figure} that, for the majority of sources, the observed break, $\Delta\Gamma$, is
systematically larger than that predicted by the EBL model, $\Delta\Gamma_{EBL}$. This is most notibly the
case for 1ES~2344+514 ($z=0.044$), PKS~2005-489 ($z=0.071$), W~Comae ($z=0.102$), PKS~2155-304 ($z=0.116$) and
H~1426+428 ($z=0.129$). The difference, $\Delta\Gamma-\Delta\Gamma_{EBL}$, is almost certainly the result of
(convex) intrinsic curvature in the spectra of these objects, which is not unexpected
\citep{2005ApJ...625..727P}, and can have several interpretations, for instance as being due to a turn-over in
the distribution of the underlying emitting particles (acceleration effects; e.g., \citealt{2006A&A...448..861M}) or to Klein-Nishina suppression
(emission effects).

Nevertheless, it is striking that there are many sources for which $\Delta\Gamma\simeq\Delta\Gamma_{EBL}$. For these sources, the intrinsic broad-band $\gamma$-ray spectra are compatible (within errors) with single power laws. For those that additionally have $\Gamma_{HE}\lesssim 2.0$, the high energy peaks are not constrained by the current observations, despite having a well defined observational $\nu F_\nu$ peak. The most striking examples are H~2356-309 ($z=0.129$), 1RXS~J101015.9-311909 ($z=0.142$), 1ES~1101-232 ($z=0.186$), 1ES~0414+009 ($z=0.287$) and S5~0716+714 ($z=0.300$).


\subsection{Constraining the EBL density}
\label{bayesian}

Since a significant fraction of the observed break can be directly attributed to the EBL, we attempt to constrain its density by applying a Bayesian model which takes into account the effects discussed in the previous section. The model is described fully in Appendix~\ref{App1}. We use two prescriptions to account for the effects of the EBL on the spectra, $\Delta\Gamma_{EBL}(E,z)$. In the first, the break is modeled as a linear function of the redshift, with coefficient $a$, \citep[as in][]{2006ApJ...652L...9S} and assume that the VHE measurements cover approximately the same energy range, so that the effect of energy threshold can be neglected.
\begin{equation}\label{EQ::DELTA_LINEAR}
\Delta\Gamma_{EBL\,i}(a) = \Delta\Gamma_{EBL}(E_i,z_i|a) \approx a z_i + O(z_i^2).
\end{equation}
In the second model we apply a scaling factor, $\alpha$, to the EBL model of Fra08, 
which results in an expected break of,
\begin{equation}\label{EQ::DELTA_FRANCESCHINI}
\Delta\Gamma_{EBL\,i}(\alpha) 
= \Delta\Gamma_{EBL}(E_i,z_i|\alpha)
= \alpha \Delta\Gamma_{Fra08}(E_i,z_i),
\end{equation}
where $\Delta\Gamma_{Fra08}(E_i,z_i)$, is calculated for each source as in
section~\ref{Prediction}.

\begin{figure*}[p]
\centering
\includegraphics[width=0.99 \textwidth]{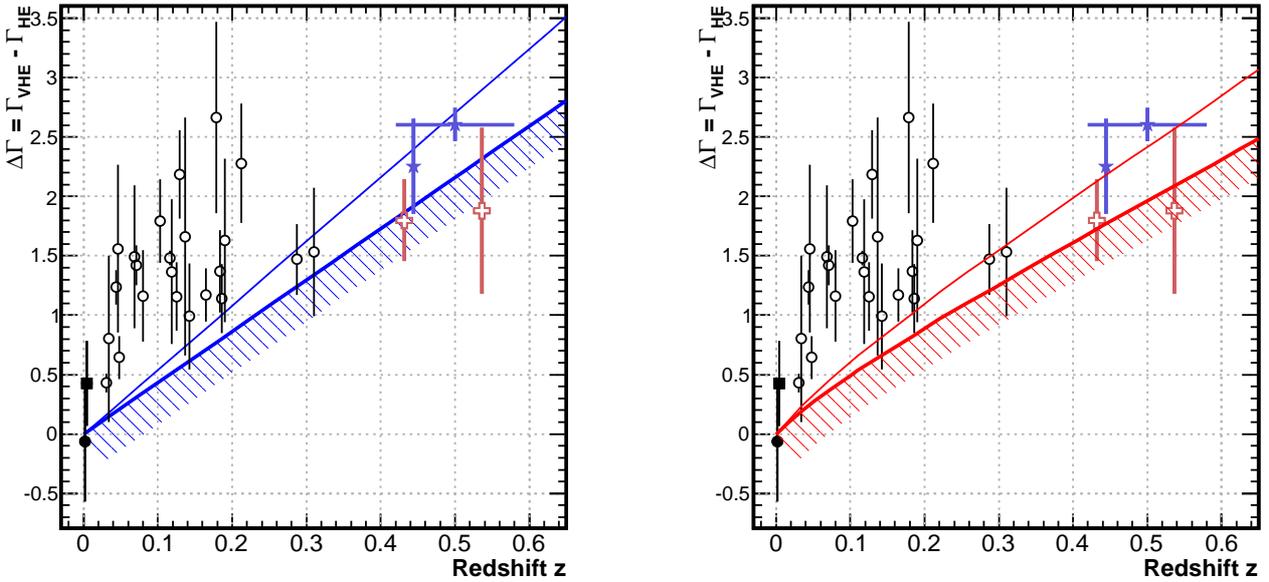}
\caption{The value of $\Delta \Gamma$ as a function of the redshift $z$. 
Overlaid are the predicted breaks obtained from the Bayesian fit (simple line), as well as the 95\% CL lower limits (hashed area). Left is the linear approximation (Equation~\ref{EQ::DELTA_LINEAR}) and right is the scaled model of Fra08 (Equation~\ref{EQ::DELTA_FRANCESCHINI}). For both models, none of the sources are significantly lower than the 95\% CL lower limit. The symbols are descibed in Figure~\ref{figure}.}
\label{figure2}
\end{figure*}

For both models, the posterior probability is computed and the results are given in Table~\ref{TAB::RESULTS_SCALED}. Figure~\ref{figure2} depicts the resulting $\Delta \Gamma$ for each model, using the mean value (i.e. $\langle a\rangle$ and $\langle \alpha\rangle$) and the 95\% confidence level (CL) lower limit (i.e. $a_{P<95\%}$ and $\alpha_{P<95\%}$).

The Bayesian model gives a value of $\langle a\rangle = 5.37 \pm 0.65$ and an 95\% CL upper limit of 6.44 for the linear EBL model, significantly less than the value of
$8.4\pm 1.0$ reported by \citet{2010A&A...522A..12Y} using a simple $\chi^2$ fit, which did not account for the intrinsic breaks. \citet{2010ApJ...709L.124S} found that their \textit{baseline model} can be approximated by a linear coefficient of $7.99$, which cannot be reconciled with the results presented here. The null hypothesis, i.e. that there is no dependence of $\Delta \Gamma$ with the redshift ($a=0$), is rejected at more than $8\sigma$. The spectral break predicted using the model of Fra08 is in good agreement with the data; the mean scaling factor is $\langle \alpha\rangle = 0.85\pm 0.10$ and the 95\% CL limit is $\alpha<1.02$.

\begin{table*}[p]
\caption{\label{TAB::RESULTS_SCALED} Summary of results with the linear parametrization and scaled \citet{2008A&A...487..837F} EBL model obtained with the Bayesian approach described in the text.}
\begin{center}\begin{tabular}{l|c|l|c}\hline\hline
Parameter & Linear model & Parameter & Scaled Fra08 model \\ \hline
Mean value: $\langle a\rangle$ & 5.37 & Mean value: $\langle\alpha\rangle$ & 0.85 \\
RMS: $\sqrt{\langle a^2\rangle - \langle a\rangle^2}$ & 0.65 & RMS: $\sqrt{\langle\alpha^2\rangle - \langle\alpha\rangle^2}$ & 0.10 \\
Upper limit: $a_{P<95\%}$ & 6.44 & Upper limit: $\alpha_{P<95\%}$ & 1.02 \\
Lower limit: $a_{P>5\%}$ & 4.32 & Lower limit: $\alpha_{P>5\%}$ & 0.69 \\\hline
\end{tabular}
\end{center}
\end{table*}

\section{Conclusions and perspectives}

We have shown that broad-band $\gamma$-ray spectra (from $\sim100$\,MeV to a few TeV) carry the imprint of the EBL and provide a unique dataset to probe its properties. The redshift dependence of the difference of the photon indices in VHE and HE range, $\Delta \Gamma$, is found to be compatible with expectations from EBL attenuation.  The Pearson correlation coefficient shows that $\Delta\Gamma$ and $z$ are significantly correlated. We developed a Bayesian model to fit the data set, accounting for intrinsic spectral softening, and find that the EBL density is consistent with the value predicted by \citet{2008A&A...487..837F}. Similar results were found by \citet{2012Sci...338.1190A} and \citet{2012arXiv1212.3409H}, who modeled the EBL-absorbed spectra of AGNs detected in the HE and VHE regimes respectively, and found scaling factors of $\alpha_{Fermi}=1.02\pm0.23$ and $\alpha_{HESS}=1.27_{-0.15}^{+0.18}$. Their approach has the potential to be more powerful than that used here, since it can probe the features of the EBL-absorption signature as a function of energy. However their approach is more reliant on the detailed modeling of the intrinsic spectra of the objects and of the EBL density and does not take advantage of the wider energy band available when the HE and VHE observations are combined. The two approaches are complimentary and yield roughly compatible results within the combined statistical and systematic errors. Taking only the statistical errors, a $\chi^2$ fit to the HESS, \textit{Fermi} and our results gives a mean value of $\alpha_{combined}\approx0.98$ and a value of $\chi^2=5.45$ for two degrees of freedom, compatible with the hypothesis that the values are consistent at the $1.85\sigma$ level. Our model also offers a simple prescription for constraining the redshift of a GeV-TeV sources based on their measured value of $\Delta\Gamma$ (see Appendix~\ref{App1}). Applying our findings to PG~1553+113 and 3C~66A leads to $z<0.64$ and $z<0.55$ respectively, in good agreement with the spectroscopic constraints.

No sources have breaks significantly smaller than $\Delta\Gamma_{EBL}$. Significant deviations from the expected EBL-induced spectral breaks could indicate either concave curvature in the intrinsic spectrum of the source or that other processes are at play during the propagation of $\gamma$ rays, such as cosmic-ray interactions along the line of sight which create spectral softening in high-redshift source spectra \citep{2012ApJ...751L..11E}. Our findings indicate that experimental uncertainties need to improve, and firmer redshift estimations established, before the significance of this effect can be assessed. 

The recent commissioning of the 28m H.E.S.S.~2 telescope, the commissioning of the upgraded MAGIC telescopes and the upgrade of the VERITAS cameras and trigger, should increase the distance at which new blazars can be detected, while the planned CTA project \citep{2011ExA....32..193A} should reduce the uncertainties mentioned above due to its superior sensitivity. Finally, a useful feature of studies of spectral breaks vs redshift, such as this one, is their capacity to provide distance estimations for BL Lacs (see, e.g., \citealt{2010ApJ...708.1310A,2010MNRAS.405L..76P,2012A&A...543A.111P}) since an estimated $50\%$ of this population has an unknown or uncertain redshift.

\bibliography{DeltaGamma}
\bibliographystyle{aa}

\appendix
\section{Bayesian model}
\label{App1}
\subsection{Development of the model}
To extract information about the EBL density from the data presented
in Figure~\ref{figure} we develop a ``hierarchical'' Bayesian model
\cite[using the terminology of][]{REF::GELMAN::BOOK2003} which is
described by source-by-source spectral parameters and global
parameters specifying, amongst other things, the EBL density of
primary interest here. We use a Bayesian methodology to write the
posterior density for the model parameters, marginalize over the
source-by-source parameters which are not of interest and produce
estimates and confidence intervals for the EBL density.

In the development below we make repeated use of the
  \textit{conditional probability rule} (CPR), that a joint probability of
  two (sets of) events $A$ and $B$, $P(A,B)$, can be expressed as
  $P(A,B)=P(A|B) P(B)$. In the case that the two events are
  independent this becomes $P(A,B)=P(A)P(B)$. We also frequently use
  the rule for \textit{marginalizing} (or integrating) over unwanted
  parameters, $P(A)=\int P(A,B) dB$. Finally, we use the standard
  identity for the product of two Gaussians \citep[see
  e.g.][]{IMM2005-03312}.  In particular if $N(x|\mu,\sigma^2)$ denotes
  a Gaussian distribution with mean $\mu$ and variance $\sigma^2$,
  then, 
\begin{multline}\label{EQ::GAUSSIANS}
\int_{-\infty}^\infty N(x|\mu_1,\sigma_1^2) N(x|\mu_2,\sigma_1^2) dx\\ 
= \frac{1}{\sqrt{2\pi}(\sigma_1^2+\sigma_2^2)^{\half}}
\exp\left\{-\frac{(\mu_1-\mu_2)^2}{2 (\sigma_1^2+\sigma_2^2)}\right\}.
\end{multline}

The data set to be modeled consists of $N$ GeV and TeV spectral
measurements, $\Gamma_{\mathrm{G}i}^M$ and $\Gamma_{\mathrm{T}i}^M$
with measurement variances of $\sigma_{\mathrm{G}i}^2$ and
$\sigma_{\mathrm{T}i}^2$ and redshifts $z_i$ (which are themselves not
considered as measurement data). In what follows we refer frequently
to the measured spectral break, $\Delta\Gamma_i^M =
\Gamma_{\mathrm{T}i}^M-\Gamma_{\mathrm{G}i}^M$. We write the data set
as $Y=\{\Gamma_{\mathrm{G}i}^M, \Gamma_{\mathrm{T}i}^M\}$.

Each source is parameterized by four values, the intrinsic spectral
indexes in the GeV and TeV regimes, $\Gamma_{\mathrm{G}i}^I$ and
$\Gamma_{\mathrm{T}i}^I$ and the spectral indexes after absorption by
the EBL, $\Gamma_{\mathrm{G}i}^A$ and $\Gamma_{\mathrm{T}i}^A$. In
what follows we refer frequently to the intrinsic spectral break,
$\Delta\Gamma_i^I =
\Gamma_{\mathrm{T}i}^I-\Gamma_{\mathrm{G}i}^I$. The global parameters
that describe the EBL absorption itself are denoted abstractly as $G$
and we write the set of all parameters of the model as
$\Theta=\{\Gamma_{\mathrm{G}i}^I, \Gamma_{\mathrm{T}i}^I,
\Gamma_{\mathrm{G}i}^A, \Gamma_{\mathrm{T}i}^A, G\}$.

Bayes' theorem allows us to write the \textit{posterior probability}
distribution of the parameters after the measurements have been made,
$P(\Theta|Y)$, in terms of the standard \textit{Likelihood} of the
data, $P(Y|\Theta)$, and the \textit{prior probability} distribution
of the model parameters, $P(\Theta)$:
\begin{displaymath}
P(\Theta|Y) \propto P(\Theta)  P(Y|\Theta)
\end{displaymath}
The relation is written as a proportionality, instead of as an
equality, as a global normalization factor has been
neglected.

The model has four primary components that we discuss below. These
are: (1) the likelihood for the GeV and TeV measurements, given the
true absorbed spectral indexes of the sources, (2) a relationship
between the absorbed index in the GeV regime and the intrinsic
(unabsorbed) index, (3) an relationship for the analogous indexes in
the TeV regime, and (4) a specification of how the intrinsic index in
the GeV regime is related to the intrinsic index in the TeV regime.

We assume the measurements of the individual indexes are independent
(no correlation between measurements from different sources or between
the GeV and TeV bands) and that the distribution for each measurement
($\Gamma_{\mathrm{G}i}^M$ or $\Gamma_{\mathrm{T}i}^M$) is Gaussian
with mean given by the appropriate absorbed index and with variance
given by the measurement errors squared. Therefore the likelihood is,
\begin{displaymath} 
P(Y|\Theta) = \prod_i N(\Gamma_{\mathrm{G}i}^M|\Gamma_{\mathrm{G}i}^A,\sigma_{\mathrm{G}i}^2) N(\Gamma_{\mathrm{T}i}^M|\Gamma_{\mathrm{T}i}^A,\sigma_{\mathrm{T}i}^2).
\end{displaymath}


Similarly for the prior, we assume that the only link between the
source-by-source index parameters for different sources comes through
the global parameters $G$. Therefore, using the CPR we can write,
\begin{displaymath} 
P(\Theta) = P(G) \prod_i P(\Gamma_{\mathrm{G}i}^I, \Gamma_{\mathrm{T}i}^I, \Gamma_{\mathrm{G}i}^A, \Gamma_{\mathrm{T}i}^A|G)
\end{displaymath}
It now remains only to describe how the intrinsic and absorbed indexes
for each source are related. We assume that the absorbed index in each
band depends only on the intrinsic index in that band and on the EBL
parameters. Repeatedly applying the CPR, this gives,
\begin{equation} \label{EQ::PRIOR_GENERAL}
P(\Gamma_{\mathrm{G}i}^I, \Gamma_{\mathrm{T}i}^I, \Gamma_{\mathrm{G}i}^A, \Gamma_{\mathrm{T}i}^A|G) = 
P(\Gamma_{\mathrm{G}i}^A|\Gamma_{\mathrm{G}i}^I, G) 
P(\Gamma_{\mathrm{T}i}^A|\Gamma_{\mathrm{T}i}^I, G)
P(\Gamma_{\mathrm{T}i}^I, \Gamma_{\mathrm{G}i}^I|G).
\end{equation} 
We further assume that there is no absorption in the GeV regime, and
that the absorption in the TeV regime changes the index in a
deterministic way. Specifically we assume that,
({\em i}) $\Gamma_{\mathrm{G}i}^A=\Gamma_{\mathrm{G}i}^I$ and 
({\em ii})
$\Gamma_{\mathrm{T}i}^A=\Gamma_{\mathrm{T}i}^I +
\Delta\Gamma_{EBL}(E_i,z_i|G)$, where $\Delta\Gamma_{EBL}(E_i,z_i|G)$
is a function giving the change in TeV index for a source at redshift
$z_i$ measured at a TeV ``threshold'' energy of $E_i$. This can be
expressed in terms of a probability using the Dirac $\delta$-function:
\begin{align*}
P(\Gamma_{\mathrm{G}i}^A|\Gamma_{\mathrm{G}i}^I, G)  &= 
\delta(\Gamma_{\mathrm{G}i}^A - \Gamma_{\mathrm{G}i}^I) &(i) \\
P(\Gamma_{\mathrm{T}i}^A|\Gamma_{\mathrm{T}i}^I, G) &=
\delta(\Gamma_{\mathrm{T}i}^A - \Gamma_{\mathrm{T}i}^I - \Delta\Gamma_{EBL}(E_i,z_i|G)) & (ii)
\end{align*}
Equation \ref{EQ::PRIOR_GENERAL} then reads:
\begin{multline*}
P(\Gamma_{\mathrm{G}i}^I, \Gamma_{\mathrm{T}i}^I,
\Gamma_{\mathrm{G}i}^A, \Gamma_{\mathrm{T}i}^A|G) \\
= \delta(\Gamma_{\mathrm{G}i}^A - \Gamma_{\mathrm{G}i}^I)
\delta(\Gamma_{\mathrm{T}i}^A - \Gamma_{\mathrm{T}i}^I - \Delta\Gamma_{EBL}(E_i,z_i|G))
P(\Gamma_{\mathrm{T}i}^I, \Gamma_{\mathrm{G}i}^I|G).
\end{multline*} 

The final and most interesting part of the model is to describe how
the intrinsic indexes in the two bands are related. We expand this
using the CPR to give,
\begin{displaymath}
P(\Gamma_{\mathrm{T}i}^I, \Gamma_{\mathrm{G}i}^I|G) =
P(\Gamma_{\mathrm{T}i}^I|\Gamma_{\mathrm{G}i}^I,G) P(\Gamma_{\mathrm{G}i}^I|G).
\end{displaymath}
We adopt a uniform prior for $\Gamma_{\mathrm{G}i}^I$,
$P(\Gamma_{\mathrm{G}i}^I)=1$, and restict ourselves to forms for the
conditional probability that can be expressed as a function of the
intrinsic break, $P(\Gamma_{\mathrm{T}i}^I|\Gamma_{\mathrm{G}i}^I,G) =
P(\Delta\Gamma_i^I|G)$, and assume that the break for each source is
drawn from a single universal distribution which has, at most, some
dependence on the global parameter set, $G$. 
The final expression for the prior for the parameters of each source is:
\begin{multline*}
P(\Gamma_{\mathrm{G}i}^I, \Gamma_{\mathrm{T}i}^I,
\Gamma_{\mathrm{G}i}^A, \Gamma_{\mathrm{T}i}^A|G) \\
= 
\delta(\Gamma_{\mathrm{G}i}^A - \Gamma_{\mathrm{G}i}^I)
\delta(\Gamma_{\mathrm{T}i}^A - \Gamma_{\mathrm{T}i}^I - \Delta\Gamma_{EBL}(E_i,z_i|G))
P(\Delta\Gamma_i^I|G).
\end{multline*}

Putting everything together, and recognizing that we haven't discussed
the parameters $G$ yet, and leaving the exact choice of prior for the
intrinsic break open for the present time, the full posterior density
is,
\begin{multline*}
P(\{\Gamma_{\mathrm{G}i}^I, \Gamma_{\mathrm{T}i}^I, \Gamma_{\mathrm{G}i}^A, \Gamma_{\mathrm{T}i}^A\},G|\{\Gamma_{\mathrm{G}i}^M, \Gamma_{\mathrm{T}i}^M\}) 
\propto\\
 P(G) \prod_i 
P(\Delta\Gamma_i^I|G)
N(\Gamma_{\mathrm{G}i}^M|\Gamma_{\mathrm{G}i}^A,\sigma_{\mathrm{G}i}^2) N(\Gamma_{\mathrm{T}i}^M|\Gamma_{\mathrm{T}i}^A,\sigma_{\mathrm{T}i}^2) \\
\delta(\Gamma_{\mathrm{G}i}^A - \Gamma_{\mathrm{G}i}^I)
\delta(\Gamma_{\mathrm{T}i}^A - \Gamma_{\mathrm{T}i}^I - \Delta\Gamma_{EBL}(E_i,z_i|G))
\end{multline*}

We marginalize over the parameters that are not of interest,
$\Gamma_{\mathrm{G}i}^I$, $\Gamma_{\mathrm{T}i}^I$,
$\Gamma_{\mathrm{G}i}^A$ and $\Gamma_{\mathrm{T}i}^A$ to give the
posterior probability for the global parameters $G$. The integrals
over the parameters for each source can be done separately,
\begin{multline*}
\mathcal{I}_i = 
\int d\Gamma_{\mathrm{G}i}^I \int d\Gamma_{\mathrm{T}i}^I
\int d\Gamma_{\mathrm{G}i}^A \int d\Gamma_{\mathrm{T}i}^A \\
P(\Delta\Gamma_i^I|G) N(\Gamma_{\mathrm{G}i}^M|\Gamma_{\mathrm{G}i}^A,\sigma_{\mathrm{G}i}^2) N(\Gamma_{\mathrm{T}i}^M|\Gamma_{\mathrm{T}i}^A,\sigma_{\mathrm{T}i}^2) \\
\delta(\Gamma_{\mathrm{G}i}^A - \Gamma_{\mathrm{G}i}^I)
\delta(\Gamma_{\mathrm{T}i}^A - \Gamma_{\mathrm{T}i}^I - \Delta\Gamma_{EBL}(E_i,z_i|G))
\end{multline*}
The integrals over $\Gamma_{\mathrm{G}i}^A$ and
$\Gamma_{\mathrm{T}i}^A$ can be done immediately against the delta
functions to give,
\begin{multline*}
\mathcal{I}_i =
\int d\Gamma_{\mathrm{G}i}^I \int d\Gamma_{\mathrm{T}i}^I
P(\Delta\Gamma_i^I|G) \\
N(\Gamma_{\mathrm{G}i}^M|\Gamma_{\mathrm{G}i}^I,\sigma_{\mathrm{G}i}^2) N(\Gamma_{\mathrm{T}i}^M|\Gamma_{\mathrm{T}i}^I+\Delta\Gamma_{EBL}(E_i,z_i|G),\sigma_{\mathrm{T}i}^2) 
\end{multline*}
Making a change of integration variable from $\Gamma_{\mathrm{T}i}^I$
to $\Delta\Gamma_i^I$, the Gaussians can be manipulated to give,
\begin{multline*}
\mathcal{I}_i =
\int d(\Delta\Gamma_i^I) P(\Delta\Gamma_i^I|G) \int d\Gamma_{\mathrm{G}i}^I\\
N(\Gamma_{\mathrm{G}i}^I|\Gamma_{\mathrm{G}i}^M,\sigma_{\mathrm{G}i}^2) 
N(\Gamma_{\mathrm{G}i}^I|\Gamma_{\mathrm{T}i}^M-\Delta\Gamma_i^I-\Delta\Gamma_{EBL}(E_i,z_i|G),\sigma_{\mathrm{T}i}^2).
\end{multline*}
The second integral can be evaluated using
equation~\ref{EQ::GAUSSIANS}. Putting all the source integrals
together gives the general expression,
\begin{multline}
\label{EQ::POSTERIOR_DGAMMA_GENERAL}
P(G|\{\Gamma_{\mathrm{G}i}^M, \Gamma_{\mathrm{T}i}^M\}) 
\propto P(G)\\\prod_i \int d(\Delta\Gamma_i^I) P(\Delta\Gamma_i^I|G)
N(\Delta\Gamma_i^I|\Delta\Gamma_i^M-\Delta\Gamma_{EBL}(E_i,z_i|G),
\sigma_{\mathrm{G}i}^2+\sigma_{\mathrm{T}i}^2).
\end{multline}

\subsection{Various priors for $\Delta\Gamma_i^I$}

We examine three concrete cases for $P(\Delta\Gamma_i^I|G)$. The first
two are based on simple assumptions and result in analytic expressions
for the full posterior probability that are easy to understand. The
final prior distribution is derived from Monte Carlo realizations of
an SSC model, as described in appendix~\ref{App2}. The results
from this final case that are presented in
section~\ref{bayesian}. Here we develop the first two cases. A
relatively weak assumption is that the TeV index can be no harder than
the GeV index. It would seem reasonable to express this using the
Heaviside step function, $\Theta(x)$, as
\begin{equation}\label{EQ::PRIOR_DGAMMA_FLAT}
P(\Delta\Gamma_i^I|G) = \Theta(\Delta\Gamma_i^I).
\end{equation}
However this is unsatisfactory, as it asserts that the mean intrinsic
break is infinite, $\langle \Delta\Gamma_i^I\rangle\rightarrow\infty$,
which is clearly not realistic. As will be seen below, this results in
an unphysical posterior distribution. To remedy this failing we
instead assume that the prior distribution of the intrinsic break is
given by a Gaussian, truncated at negative values:
\begin{equation}\label{EQ::PRIOR_DGAMMA_GAUSSIAN}
P(\Delta\Gamma_i^I|G) = \Theta(\Delta\Gamma_i^I)
N(\Delta\Gamma_i^I|\mu_I,\sigma_I^2).
\end{equation}
In this case the prior is parameterized by two values, $\mu_I$ and
$\sigma_I$, which must be either estimated in the problem (i.e.\ added
to the global parameter set $G$), or specified externally. We will
simply assume $\mu_I=0$ and derive the results for various reasonable
values of $\sigma_I$.

Equation~\ref{EQ::POSTERIOR_DGAMMA_GENERAL} can be used to calculate
the posterior distribution in the three cases for
$P(\Delta\Gamma_i^I)$ discussed above. In the second case, with the
prior given by Equation~\ref{EQ::PRIOR_DGAMMA_GAUSSIAN}, the final
integration can be done to give, after applying the formula for the
product of Gaussians,
\begin{multline}\label{EQ::POSTERIOR_DGAMMA_GAUSSIAN}
P(G|\{\Gamma_{\mathrm{G}i}^M, \Gamma_{\mathrm{T}i}^M\},\mu_I=0,\sigma_I) 
\propto \\P(G)\prod_i \exp\left\{-\frac{(\Delta\Gamma_{EBL}(E_i,z_i|G)-\Delta\Gamma_i^M)^2}
{2 (\sigma_\mathrm{G_i}^2+\sigma_\mathrm{T_i}^2+\sigma_I^2)}\right\}\\
\mathrm{erfc}\left\{\frac{\sigma_I(\Delta\Gamma_{EBL}(E_i,z_i|G)-\Delta\Gamma_i^M)}
{\sqrt{2}(\sigma_\mathrm{G_i}^2+\sigma_\mathrm{T_i}^2)^\half
(\sigma_\mathrm{G_i}^2+\sigma_\mathrm{T_i}^2+\sigma_I^2)^\half}
\right\},
\end{multline}
where $\mathrm{erfc}(x)=\frac{2}{\sqrt{\pi}}\int_x^\infty dx^\prime\,e^{-x^{\prime2}}$ 
is the complementary error function.

It is instructive to examine the two limiting cases of $\sigma_I=0$
and $\sigma_I\rightarrow\infty$. In the first case we arrive at,
\begin{multline*}
\log P(G|\{\Gamma_{\mathrm{G}i}^M,
\Gamma_{\mathrm{T}i}^M\},\mu_I=0,\sigma_I=0) \\
\propto \sum_i \frac{(\Delta\Gamma_{EBL}(E_i,z_i|G)-\Delta\Gamma_i^M)^2}
{\sigma_\mathrm{G_i}^2+\sigma_\mathrm{T_i}^2},
\end{multline*}
which is exactly the expression that would result from a simple
least-squares fit to the measured spectral breaks. In the second case
($\sigma_I\rightarrow\infty$) we have,
\begin{multline}\label{EQ::POSTERIOR_DGAMMA_FLAT}
P(G|\{\Gamma_{\mathrm{G}i}^M,
\Gamma_{\mathrm{T}i}^M\},\mu_I=0,\sigma_I\rightarrow\infty) \\
\propto \prod_i \mathrm{erfc}\left\{\frac{\Delta\Gamma_{EBL}(E_i,z_i|G) - \Delta\Gamma_i^M}
{\sqrt{2}(\sigma_{\mathrm{G}i}^2+\sigma_{\mathrm{T}i}^2)^{\half}}\right\},
\end{multline}
which is the same expression as would be derived starting from the
prior given by Equation~\ref{EQ::PRIOR_DGAMMA_FLAT}. We therefore have
an expression that transforms continuously between the two clearly
identifiable extremities as a function of $\sigma_I$.

As described in section~\ref{bayesian}, we attempt to derive
constraints on two simple models for the EBL. First, a linear
approximation, given by Equation~\ref{EQ::DELTA_LINEAR}, and second, a
scaling of the EBL model of \citet{2008A&A...487..837F}, as described
by Equation~\ref{EQ::DELTA_FRANCESCHINI}. In both cases we assume a
uniform prior in the positive region of the parameter space for $a$,
$P(a)=\Theta(a)$, and $\alpha$, $P(\alpha)=\Theta(\alpha)$,
respectively.

Using either of these, the deficiencies in the model given by
Equation~\ref{EQ::PRIOR_DGAMMA_FLAT} is finally evident. Combining
Equations~\ref{EQ::DELTA_FRANCESCHINI} and
\ref{EQ::POSTERIOR_DGAMMA_FLAT} we get,
\begin{multline}
P(\alpha|\{\Gamma_{\mathrm{G}i}^M,
\Gamma_{\mathrm{T}i}^M\},\mu_I=0,\sigma_I\rightarrow\infty) 
\propto \\ \Theta(\alpha) \prod_i \mathrm{erfc}\left\{\frac{\alpha\Delta\Gamma_{Fra08}(E_i,z_i) - \Delta\Gamma_i^M}
{\sqrt{2}(\sigma_{\mathrm{G}i}^2+\sigma_{\mathrm{T}i}^2)^{\half}}\right\}.
\end{multline}
The most probable value occurs at $\alpha=0$, which is consistent with the assertion in this case that $\langle\Delta\Gamma_i^I\rangle\rightarrow\infty$.

The results in section~\ref{bayesian} are derived from combining
Equations~\ref{EQ::DELTA_LINEAR} (or \ref{EQ::DELTA_FRANCESCHINI}) and
\ref{EQ::POSTERIOR_DGAMMA_GENERAL} with the prior calculated in
appendix~\ref{App2} and integrating numerically.

\subsection{Results with different priors}

Table \ref{TAB::RESULTS_PRIO} presents the results obtained with the
half-Gaussian prior of Equation~\ref{EQ::PRIOR_DGAMMA_GAUSSIAN} (with
$\mu_I=0$), for four values for the variance of the distribution of
the intrinsic break, $\sigma_I=0.25,0.5,1.0\ \mathrm{and}\ 2.0$, and
using the prior derived from SSC modeling, illustrated in
Figure~\ref{AppSSCfig1}.

With increasing value of $\sigma_I$, the most probable value of $a$ or
$\alpha$ decreases, since a lower EBL-induced break is needed to
reproduce the data. Results derived with values of $\sigma_I=0.5\
\mathrm{and}\ 1.0$ are in good agreement with those from the SSC
model.

\begin{table*}[p]
\caption{\label{TAB::RESULTS_PRIO} Summary of results from Bayesian
  model with different priors.}
\centering 
\begin{tabular}{l|cccc|c}\hline\hline
& \multicolumn{4}{c|}{Half-Gaussian prior 
(Equation~\ref{EQ::PRIOR_DGAMMA_GAUSSIAN}, $\mu_I=0$)} & SSC prior \\
Parameter & $\sigma_I=0.25$ & $\sigma_I=0.5$ & $\sigma_I=1$ & $\sigma_I=2.0$ &
(Figure~\ref{AppSSCfig1})  \\ \hline
Mean value: $\langle a\rangle$ & 7.40 & 6.04 & 4.64 & 3.12 & 5.37 \\
RMS: $\sqrt{\langle a^2\rangle - \langle a\rangle^2}$ & 0.55 & 0.64 &
0.84 & 1.24 & 0.65\\
Upper limit: $a_{P<95\%}$ & 8.29 & 7.08 & 5.97 & 5.02 & 6.4 \\
Lower limit: $a_{P>5\%}$ & 6.5 & 4.99 & 3.20 & 0.9 & 4.32 \\\hline

Mean value: $\langle \alpha\rangle$ & 1.16 & 0.96 & 0.74 & 0.50 & 0.85 
\\
RMS: $\sqrt{\langle \alpha^2\rangle - \langle \alpha\rangle^2}$ & 0.08 &
0.10 & 0.13 & 0.20 & 0.10 \\
Upper limit: $\alpha_{P<95\%}$ & 1.31 & 1.12 & 0.95 & 0.80 & 1.02 \\
Lower limit: $\alpha_{P>5\%}$& 1.03 & 0.80 & 0.52 & 0.15 & 0.69 \\\hline
\end{tabular}
\end{table*}

\subsection{Constraints on the redshift}

The Bayesian methodology can also be used to constrain the
  redshifts of GeV-TeV blazars from their measured spectral breaks. In
  the case of a single source,
  equations~\ref{EQ::POSTERIOR_DGAMMA_GENERAL} and
  \ref{EQ::DELTA_FRANCESCHINI} can be adopted to express the posterior
  probability for the parameters $G=\{\alpha, z\}$, given their priors
  $P(\alpha)$ and $P(z)$\footnote{We neglect the dependence of the
    prior for $z$ on the strength of the EBL ($\alpha$), i.e. we
    assume incorrectly that $P(z|\alpha)=P(z)$.}. This can then be
  marginalized over $\alpha$ to give,
\begin{multline}
P(z|\Gamma^M_G, \Gamma^M_T) \propto P(z) \int d\alpha P(\alpha)
\int d(\Delta\Gamma^I) P(\Delta\Gamma^I) \times\\
N(\Delta\Gamma^I|\Delta\Gamma^M-\alpha\Delta\Gamma_{Fra08}(E,z|G),
\sigma_\mathrm{G}^2+\sigma_\mathrm{T}^2).
\end{multline}
The prior for the redshift, $P(z)$, could be estimated from
  the redshift distribution of detected GeV-TeV detected blazars,
  i.e.\ from the values presented in Table~\ref{table:list}. However
  the true distribution is probably not well represented by the small
  number of sources detected, so we seek an alternative approach.
  Another option is to use a flat distribution, which is conservative
  but also unrealistic. We instead compromise and use the distribution
  of 2FGL BL Lac objects and unknown AGN. Since Fermi AGN are detected
  to higher redshifts than those at TeV energies, this should still be
  a conservative approach. For the prior on the EBL scaling,
  $P(\alpha)$, we use the positive portion of a Gaussian with mean 1.0
  and RMS 0.3, $P(\alpha)=\Theta(\alpha)N(\alpha|1.0,0.3^2)$. Finally,
  we use the prior on $\Delta\Gamma^I$ derived from our SSC
  simulations, as described above.

The mean redshift and upper limits for some values of
  $\Delta\Gamma$, calculated using a typical value of
  $\sigma_\mathrm{G}^2+\sigma_\mathrm{T}^2=0.2^2$, are given in
  Table~\ref{TAB::PREDIC_Z}.  The corresponding relation between
  $\langle z\rangle$ and $\Delta\Gamma$ can be approximated
  by $$\langle z\rangle \approx 0.024 + 0.079\Delta\Gamma+
  0.022\Delta\Gamma^2 - 0.0010\Delta\Gamma^3$$ and the relation
  between $z_{P<95\%}$ and $\Delta\Gamma$ by $$z_{P<95\%} \approx
  0.081 + 0.081\Delta\Gamma+ 0.080\Delta\Gamma^2 -
  0.011\Delta\Gamma^3.$$

\begin{table*}[p]
\caption{\label{TAB::PREDIC_Z} Prediction of mean redshift value $\langle z\rangle$ and upper limit $z_{P<95\%}$ for a give value of $\Delta\Gamma$.}
\centering 
\begin{tabular}{ccc}\hline\hline
$\Delta\Gamma$ & $\langle z\rangle$ & $z_{P<95\%}$  \\ \hline
0.5  &  0.07  &  0.14\\
1.0  &  0.12  &  0.24\\
1.5  &  0.19   & 0.34\\
2.0   & 0.26  &  0.46\\
2.5 &   0.34  &  0.60\\
3.0  &  0.43  &  0.75\\
3.5  &  0.53  &  0.86\\
4.0   & 0.62  &  0.95\\\hline
\end{tabular}
\end{table*}

\section{Synchrotron Self-Compton Simulations}

\subsection{Determination of the intrinsic break properties}
\label{App2}

In order to derive a plausible prior probability density for the
intrinsic break between HE and VHE, for use in the Bayesian model, we
produce a set of Monte Carlo simulations of hypothetical BL Lacs using a
one-zone synchrotron self-Compton (SSC) model \citep{THEO::SSC_BAND},
which is often used to successfully reproduce the time-average SED from
radio to TeV energies.

We assume a spherical emission zone, with a size $R$, moving at a bulk
Doppler factor $\delta$. This region is filled by a uniform magnetic
field $B$, and a population of electrons with a density $N_e(\gamma)$
is responsible for the synchrotron emission. The synchrotron photons
are upscattered by the same population of electrons to produce
$\gamma$ rays.

The distribution of electrons is described by a power-law with an
exponential cut-off \citep{2012ApJ...753..176L},
$N_e(\gamma)=N_0\gamma^p\cdot\exp(-\gamma/\gamma_{\rm cut})$. The
model therefore has three parameters to describe the electron
population ($N_0$, $p$ and $\gamma_{\rm cut}$) and four to describe
the jet properties ($z$, $R$, $\delta$ and $B$). Among these
parameters, $R$, $\delta$ and $N_0$ have only an achromatic effect,
and $z$ produces a small K correction, which is evaluated separately
in appendix~\ref{App3}. The spectral break is also insensitive to the
value of the index $p$ of the electron distribution. The parameters
which determine the intrinsic break are therefore $B$ and 
$\gamma_{\rm  cut}$.

We perform $10^5$ simulations in which the values of $B$ and
$\gamma_{\rm cut}$ and uniformly drawn in the range $0.01<B<0.5$\,G
and $3\times 10^4<\gamma_{\rm cut}<1\times 10^7$. The other parameters
are kept fixed at the values given in Table~\ref{Apen:tableSSC}, which
are typical for BL~Lacs.

\begin{table}[p]
\caption{Parameters used for the SSC simulations.}
\label{Apen:tableSSC} 
\centering 
\begin{tabular}{c  c } 
\hline\hline 
Parameters & Value     \\ 
\hline 
$z$ & 0.1  \\
$R$ & $4.5\times 10^{16}$\,cm   \\
$\delta$ & $20$  \\
\hline 
$p$ & $2.3$  \\
$N_{0}$ & $3\times 10^3$\,cm$^{-3}$ \\
\hline\hline 
$B$ & $0.01-0.5$\,G  \\
$\gamma_{\rm cut}$ & $3\times 10^4-1\times 10^7$\\
\hline 
\end{tabular}
\end{table}

\begin{figure*}[p]
\centering
\includegraphics[width=0.99 \textwidth]{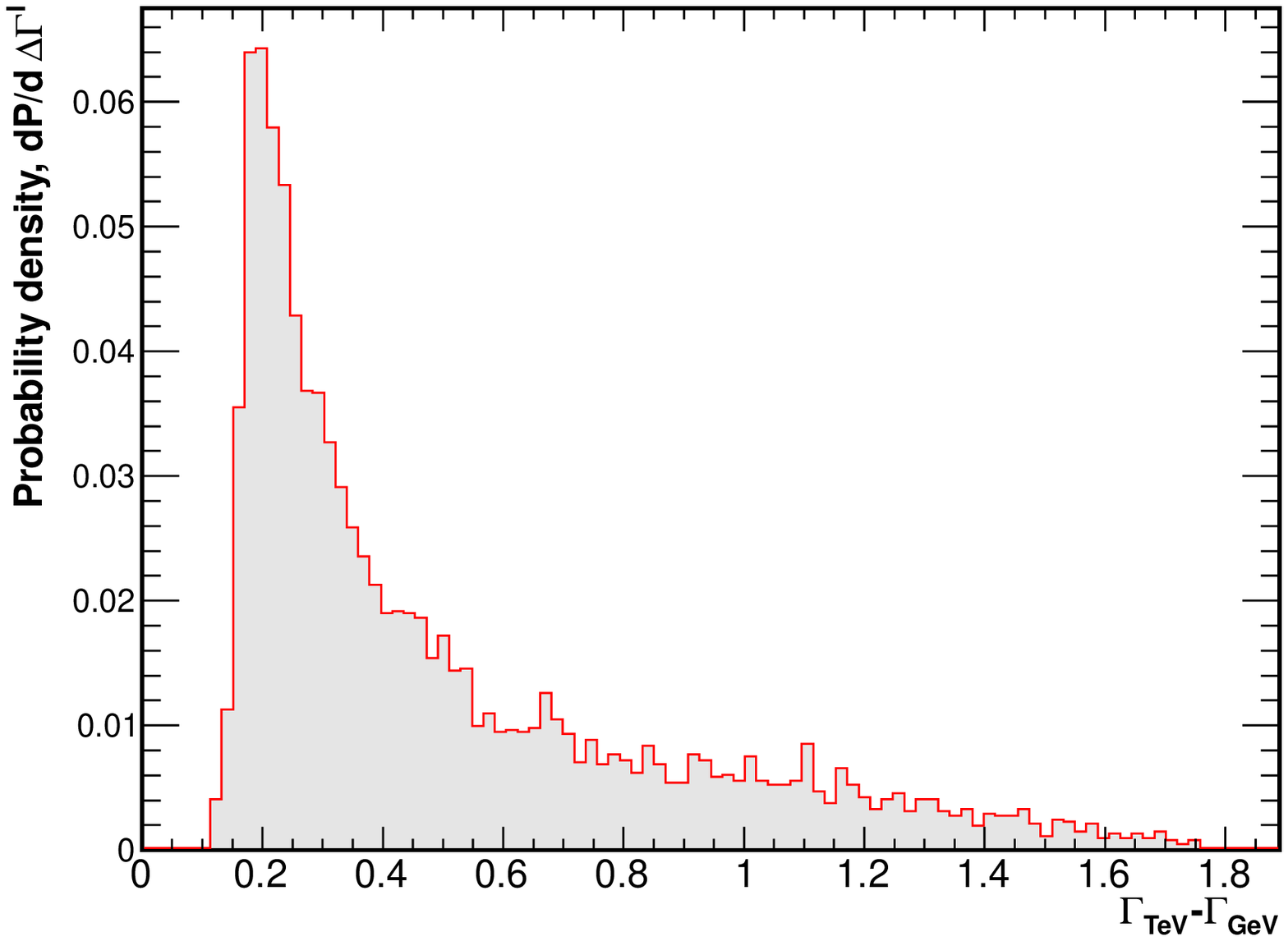}
\caption{$\Gamma_{\rm VHE}-\Gamma_{\rm HE}$ distribution (red
  histogram) obtained with the SSC simulations and application of cuts
  described in the text.}
\label{AppSSCfig1}
\end{figure*}

Since only the BL~Lac-type SEDs are of interest in this study,
simulations for which the synchrotron emission peaks at energies lower
than 10~eV or higher that 10 MeV and having a HE index of $\Gamma>2$
have been removed. The distribution of the intrinsic break
$\Gamma^I_{\rm VHE}-\Gamma^I_{\rm HE}$, depicted in Figure
\ref{AppSSCfig1}, has a sharp rise below 0.2 and a long tail at higher
break values.

\subsection{Effects of the energy shift due to the distance}
\label{App3}

SSC simulations can be used to evaluate the size of the K correction
required to account for the redshifting of the intrinsic spectrum into
the fixed HE and VHE observation windows. This effect produces a trend
of increasing observed $\Delta\Gamma$ with $z$, even in the absence of
EBL absorption.

\begin{figure*}[p]
\centering
\includegraphics[width=0.99 \textwidth]{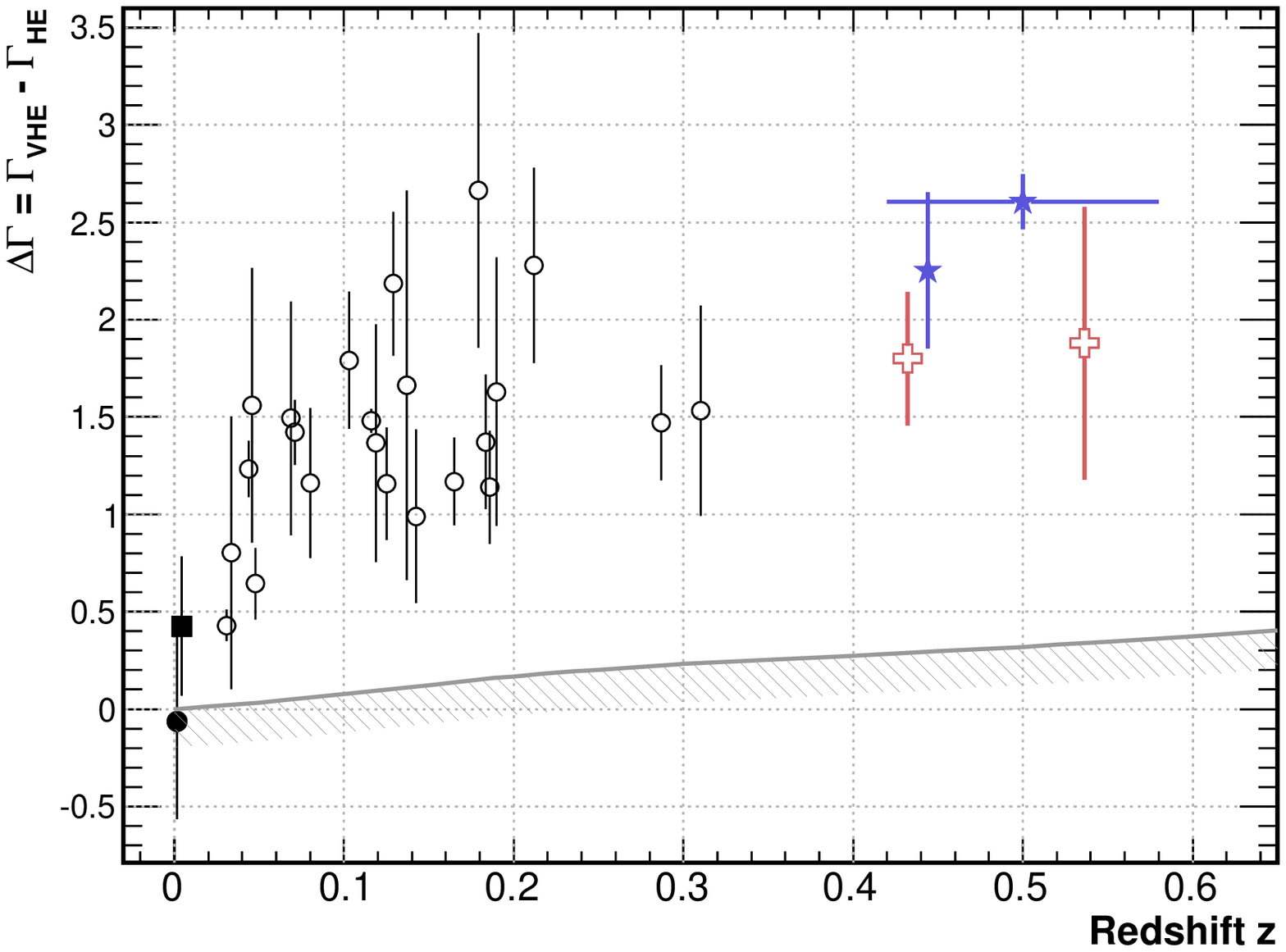}
\caption{The value of $\Delta \Gamma$ as a function of the redshift $z$. The grey line is the break due to redshift effect as computed with the SSC simulation.}
\label{AppSSCfig2}
\end{figure*}

As before, the SSC parameters are fixed to the values in
Table~\ref{Apen:tableSSC}, but with $B=0.1\,\rm{G}$ and $\gamma_{\rm cut}=
1.6\times 10^5$. This produces an intrinsic spectrum with
$\Gamma^I_{\rm HE}=1.85$ and $\Gamma^I_{\rm VHE}=3.04$,
i.e. $\Delta\Gamma_I=1.19$. Simulating the same source with redshifts
in the range $10^{-4}<z<0.7$, and with no EBL absorption, leads to an
additional component in the observed $\Delta\Gamma$ which increases
with redshift, as depicted in Figure~\ref{AppSSCfig2}. This
corresponds to the K correction for this intrinsic spectral shape and
is too small to explain the trend observed in the data.

\end{document}